\documentclass{optica-article}

\journal{opticajournal} 

\articletype{Research Article}

\usepackage{physics}

\begin{document}

\title{Device-Independent Quantum Key Distribution with realistic single-photon source implementations}

\author{Eva M. González-Ruiz,\authormark{1,2,*,$\dagger$} Javier Rivera-Dean,\authormark{2,*,$\S$} Marina F. B. Cenni, \authormark{2} Anders S. Sørensen, \authormark{1} Antonio Acín, \authormark{2,3} and Enky Oudot\authormark{2}}

\address{\authormark{1}Center for Hybrid Quantum Networks (Hy-Q), Niels Bohr Institute\\
University of Copenhagen, Blegdamsvej 17, DK-2100 Copenhagen, Denmark\\
\authormark{2}ICFO - Institut de Ciencies Fotoniques, The Barcelona Institute
of Science and Technology, 08860 Castelldefels (Barcelona), Spain\\
\authormark{3}ICREA - Instituci\'o Catalana de Recerca i Estudis Avan\c cats, Llu\'{\i}s Companys 23, 08010 Barcelona, Spain\\
\authormark{*}The authors contributed equally to this work.}

\email{\authormark{$\dagger$}eva.ruiz@nbi.ku.dk}
\email{\authormark{$\S$}javier.rivera@icfo.eu}


\begin{abstract*}
  Device-Independent Quantum Key Distribution (DIQKD) aims at generating secret keys between distant parties without the parties trusting their devices. 
  We investigate a proposal for performing fully photonic DIQKD, based on single photon sources and heralding measurements at a central station placed between the two parties. We derive conditions to attain non-zero secret-key rates in terms of the photon efficiency, indistinguishability and the second order autocorrelation function of the single-photon sources. Exploiting new results on the security bound of such protocols allows us to reduce the requirements on the physical parameters of the setup. Our analysis shows that in the considered schemes, key rates of several hundreds of secret bits per second are within reach at distances of several tens of kilometers.
\end{abstract*}

\section{Introduction}
The use of quantum mechanics 
to establish secure communication between two parties, Alice and Bob, 
is the essence of Quantum Key Distribution (QKD) where the parties use a quantum channel to establish a secret key \cite{BB84,Ekert91}. Experimental demonstrations of QKD protocols have been successfully carried out in a 
number of setups, see for instance Ref.~\cite{Scarani09}, and QKD devices are now commercially available. The security of QKD protocols relies on several assumptions, in particular: 
\begin{enumerate}
    \item Quantum mechanics is a valid theory,
    \item Alice and Bob trust their device and therefore assume that they are performing the operations described by the QKD protocol.
\end{enumerate}

 The aim of Device-Independent Quantum Key Distribution (DIQKD) \cite{Acin07} is to remove the last assumption, which can be a source of various possible hacking attacks \cite{Lydersen10}. The security of a DIQKD protocol depends on the statistics obtained by the parties without requiring any physical model of the measurements or of the state used in the protocol. The security of such protocols has been successfully proven even when allowing an eavesdropper, Eve, to perform general attacks, see e.g. Ref.~\cite{Friedman19}.
 
A DIQKD protocol is an entanglement based protocol where Alice and Bob possess two measurement apparatus, which are considered to be black boxes, on which one can define \emph{settings} and obtain \emph{outcomes}. The security of the protocol relies on the violation of a Bell inequality \cite{Acin07}. Intuitively, for some Bell inequalities, e.g. the Clauser-Horne-Shimony-Holt (CHSH) inequality \cite{Clauser69}, if the two parties observe a maximal violation, they share a maximally entangled two-qubit state \cite{mayers_self_2004,Kaniewski16} and thus cannot be correlated with a third party. In practice, however, the maximum violation of such an inequality can never be achieved because of the inevitable noise present in experiments. Reduced violations introduce the possibility that the two parties are correlated with Eve, so that when the Bell violation becomes smaller than a critical value, no secret key can be established. Unfortunately the noise requirements needed to produce a secret key are very challenging to meet in practice. 

The implementation of standard photonic-based QKD is based on the distribution of entangled photons. However, the high requirements with respect to noise make direct transmission of an entangled pair of photons undesirable for DIQKD. Indeed the unavoidable transmission losses between the two parties prevent such scheme from working unless losses are very small, that is, the users are very close. To circumvent transmission losses one can use a heralding scheme, where entanglement is generated between the local stations conditioned on the detection of photons at a central heralding station (CHS), which performs a Bell state measurement (BSM) on photonic fields entangled with each of the two stations.
This is a major advantage since successful entanglement is heralded. Photon loss thus does not influence the quality of the entangled states; it lowers the success probability but the protocol remains secure.

The first experimental demonstrations of DIQKD have been performed very recently using heralding schemes \cite{nadlinger_experimental_2022,Lim21}. In those experiments the state at the local stations are encoded in  
material qubits, in particular Ref.~\cite{nadlinger_experimental_2022} uses two trapped ions and Ref.~\cite{Lim21} uses two trapped rubidium atoms. The advantage of such encoding lies in the very high efficiency of the measurement of the matter degree of freedom, allowing for a sufficiently high violation of the CHSH inequality to perform DIQKD. On the other hand, such encodings into material qubits typically require rather advanced setups and often suffer from low repetition rates, making the application of such setups less desirable for practical applications. In fact, the experiment of Ref.~\cite{nadlinger_experimental_2022} was able to report a positive rate between users distant by only 2 meters, while the experiment of Ref.~\cite{Lim21} demonstrated a Bell violation between users distant 400 meters that was large enough in principle to establish a secure key, although the number of runs in the experiment were not sufficient to achieve this.

Photonic experiments can in principle alleviate these problems because they have a higher repetition rate, but suffer from limited measurement efficiency. Heralding schemes based on photonic implementations using a spontaneous parametric down conversion (SPDC) source and a qubit amplifier \cite{Sangouard10} have been proposed as a means to perform DIQKD, but suffer from an unfavorable scaling with distance, see e.g Appendix B of  Ref.~\cite{main_article}. On the other hand, single-photon sources, and in particular those based on InGaAs quantum dots, are able to generate on-demand single photons with high indistinguishability and purity, up to 96\% and 99.4\% respectively \cite{Tomm:2021vq,Uppueabc8268}. These single-photon sources are currently being developed towards fully integrated devices, and are thus suitable candidates for scalability of DIQKD implementations in the near future. Recently, heralding schemes for DIQKD based on such single-photon sources \cite{main_article} have been developed and these were shown to have a more promising scaling with distance, although requiring high detection efficiencies.

The security of the experimental protocols presented above was based on the violation of the CHSH inequality. Such security proofs of DIQKD have been extended to generalized CHSH inequalities \cite{Sekatski21,Woodhead21}. The choice of the correct Bell inequality is crucial, as different inequalities can lead to higher robustness of the protocol with respect to noise. Recently, a breakthrough regarding the choice of inequality was achieved in Refs.~\cite{brown_device-independent_2021,brown_device-independent_2021_arxiv}, where the authors derived a method for guaranteeing the security of protocols based on all the observed statistics. This method is equivalent to the optimization of the security of the protocol with respect to all possible Bell inequalities. This approach thus allows for improving the noise robustness of DIQKD protocols, and has shown to be useful as for developing new efficient semidefinite programming (SDP) hierarchies for device-dependent QKD protocols \cite{araujo2022}.

Over the past 20 years, there have been impressive advancements in the tools used for deriving security proofs. These advancements have led to a significant improvement in the requirements for experimental parameters. However, we are approaching the point where further improvements may be limited, as explicit eavesdropping attacks have demonstrated parameters that closely approach the established security bound \cite{farkas_bell_2021}. Consequently, one needs to find a platform which can live up to the requirements. In particular, the number of successful experimental events observed in recent loophole-free Bell test experiments is prohibitively low for long-distance DIQKD \cite{tan_improved_2020}. In this study, we address this concern by utilizing the latest development in theoretical tools to improve the experimental requirements for a purely photonic implementation of DIQKD with quantum dots single-photon sources, as proposed in Ref.~\cite{main_article}. Specifically, we conduct a finite-size analysis, which demonstrates that this platform is suitable for performing DIQKD at medium range. We improve the previously derived experimental requirements for the purely photonic implementation of DIQKD with single-photon sources proposed in Ref.~\cite{main_article}, and further analyzed in Ref.~\cite{eva2021bell}. The main contribution of this work is to study the experimental requirements for photonic DIQKD experiments with single-photon sources.

The article is organized as follows: in Section~\ref{Sec:Setup}, we present the proposed setup and characterize the main parameters that limit the experimental implementation. In Section~\ref{Sec:DIQKD}, we briefly review the security of the protocol and the method used in Refs.~\cite{brown_device-independent_2021,brown_device-independent_2021_arxiv}. In Section~\ref{Sec:Results}, we analyze the robustness of the protocol in a realistic scenario where we consider limited local efficiency, indistinguishability and purity of the single-photon source. Here, we also perform a finite statistics analysis of our results and study the security of the protocol with respect to time and distance.

\begin{figure}
\centering
\includegraphics[width=\linewidth]{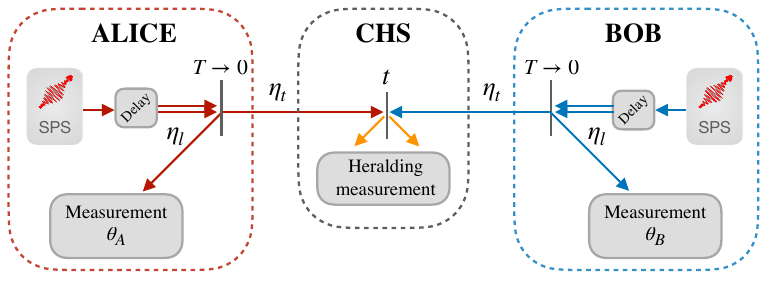}
\caption{\label{fig:scheme} Heralding set-up introduced in Ref. \cite{main_article}. Alice and Bob each possess a single-photon source, which by means of an optical delay circuit provides two simultaneous photons with orthogonal polarizations. One of them is transmitted from each lab through a low transmittance beam splitter ($T\rightarrow0$) with efficiency $\eta_t$ to a central heralded station (CHS) that performs a Bell state measurement. The entanglement of the polarization state of Alice and Bob's remaining photons after the CHS measurement can be tuned depending on the transmittance $t$ of the beam splitter at the CHS. The basis choice of the local measurements is defined by the angle $\theta_A$ ($\theta_B$) of a combination of half-wave and quarter-wave plates \cite{eva2021bell}. 
The total local efficiency at Alice and Bob's stations is denoted by $\eta_l$. }
\label{Fig:CHS:setup}
\end{figure}

\section{Experimental setup}\label{Sec:Setup}

In this work, we consider the set-up that is pictorially presented in Fig.~\ref{Fig:CHS:setup}, where Alice and Bob generate two single-photons, each with orthogonal polarization, by means of a single-photon source and an optical delay line \cite{eva2021bell}. The photons are sent towards a low-transmittance ($T\rightarrow0$) beam-splitter such that, to a very good approximation, the detection of two photons at the CHS signals that at each user's local station, one of the photons was sent to the CHS and the other is kept. Thus, the outcome of the measurement at the CHS heralds the successful generation of the desired entangled polarization state between the remaining photons at Alice and Bob's stations. In this way, the expected low transmission efficiency between Alice and Bob due to their spatial separation does not open the detection loophole; since photons at the local stations have propagated a short distance, they have only suffered limited loss, and are thus detected with very high probability.

\begin{figure}
\centering
\includegraphics[width=4cm]{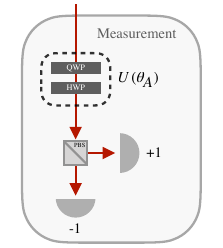}
\caption{\label{fig:scheme_measurement}
Measurement set-up in Alice's station. The half-wave plate (HWP) and quarter-wave plate (QWP) set the basis choice by fixing the relative angle $\theta_A$. A polarising beam splitter (PBS) directs photons with orthogonal polarisation to two detectors, whose outcomes are identified as $\pm 1$ respectively. 
}
\label{Fig:measurement:setup}
\end{figure}

The heralding signal at the CHS creates an entangled state in polarization by performing a Bell state measurement
\cite{BSM}. The entangled state is proportional to
\begin{equation}
 \ket{\psi} \propto \ket{+-} - \frac{t}{1-t}\ket{-+}\,,
\end{equation}
where $\ket{\pm} = \frac{1}{\sqrt{2}}\left(\ket{H}\pm\ket{V} \right)$, with $\ket{H}$ and $\ket{V}$ the states with horizontal and vertical polarization, respectively, and $t$ is the transmittance of the beam splitter at the CHS.
Note that by varying $t$, the state shared by Alice and Bob can vary from a Bell state to a non-maximally entangled state arbitrary close to a separable state. If $t=0.5$, the state corresponds to the Bell state $\ket{\psi^{-}} = \frac{1}{\sqrt{2}}\left(\ket{HV}-\ket{VH}\right)$ and we recover the set-up and results from Ref.~\cite{eva2021bell}, while if $t=0$ or 1 no entanglement is created. Being able to optimize the degree of entanglement in non-maximally entangled states is a very useful feature, as it can reduce the required efficiency for the violation of Bell's inequality \cite{eberhard}. 
Crucially, with the presented scheme the state shared between the two parties does not depend on the channel transmission efficiency $\eta_t$, and therefore allows for extending the communication distance without compromising the violation of a Bell inequality and, consequently, the security of the protocol. The local efficiency $\eta_l$, defined as the total efficiency between the source and the measurement devices (including the detection and source's efficiencies) at Alice and Bob's stations, remains however as the main challenge since this measurement is not heralded. Furthermore, due to the different experimental errors (limited indistinguishability, purity and non number-resolving detectors) the state is not a pure state.

The local measurements of Alice and Bob are performed in the polarization basis. Before detecting the photons, they transform the state according to a linear transformation $U(\theta_A)\otimes U(\theta_B)$ by means of a set of half-wave plates and quarter-wave plates (see Fig.~\ref{Fig:measurement:setup}). Once the state has been successfully rotated to the desired basis, the photons are directed to a polarising beam splitter that sends them to two different detectors. Thereby they can measure the state in any desired basis. Note that one can find a plane in the Bloch sphere such that only the relative angle between the half-wave and quarter-wave plates (which we define as $\theta_A$ and $\theta_B$ for Alice's and Bob's measurement, respectively) is the degree of freedom to optimise, which simplifies greatly the computation later on.

We consider non photon-number resolving detectors, that is, they only measure the arrival of photons. Thus the detectors are defined by the Positive-Operator-Valued-Measure (POVM) operators $\{\hat{M}_{j,A}\}$ ($\{\hat{M}_{j,B}\}$) with $j\in \{0,1\}$, where $\sum_j {\hat{M}_{j,A}}= \mathbb{I}$ ($\sum_j {\hat{M}_{j,B}}= \mathbb{I}$) and 
\begin{align}
\begin{split}
        \hat{M}_{0,{A}} = \ket{\emptyset}\bra{\emptyset}_{A}, \quad \hat{M}_{1,{A}} &= \mathbb{I} - \ket{\emptyset}\bra{\emptyset}_{A}\,,
        \label{eq:POVM}
\end{split}
\end{align}
and similarly for Bob. 
The state $\ket{\emptyset}$ is the vacuum state, and the subscripts $1$ and $0$ define the different outcomes that can be obtained. Depending on which detector clicked, they obtain a classical outcome $\pm1$. If no detector clicked ($\emptyset$), they assign the outcome to $+1$. This is a choice that has been shown to perform as the best achievable strategy \cite{eva2021bell,eta8284}. Since the detectors are not number-resolving, this induces errors for certain detection combinations. For instance, the two photons generated by Alice may be reflected at the first beam splitter, while both of Bob's photons are transmitted to the CHS giving the correct detection pattern. In principle, detection events consisting of two photons coming from the same station are suppressed by the pair of half-wave plates placed in the transmission path \cite{main_article}. However, due to the imperfect indistinguishability of the sources this suppression is not completely effective (see Ref. \cite{eva2021bell} for details).


We denote by $\hat{\rho}$ the state of Alice and Bob after a heralding event at the CHS. The probability of detection $p(a_i,b_j|x_d,y_q)$ has classical inputs $\{x_d,y_q\}$ and classical outputs $\{a_i,b_j\}$, where $x_d = \theta_{A,d}$ ($y_q = \theta_{B,q}$) represents Alice's (Bob's) input. Thus the probability of detection is given by 
\begin{equation}
    p(a_i,b_j|x_d,y_q) = \Tr{\left(\hat{M}_{i,{A}}\otimes\hat{M}_{j,{B}}\right) \hat{\rho}'}\, ,
\end{equation}
where $\hat{\rho}'=U(\theta_{A,d})\otimes U(\theta_{B,q})\hat{\rho} U^{\dagger}(\theta_{A,d})\otimes U^\dagger(\theta_{B,q})$. 
Each of these probability contributions includes the probability of all the events that lead to the acceptance at the CHS and not only the ideal scenario (see Table S1 in the Supplementary Material). Following Ref.~\cite{eva2021bell} we label each of these probability terms as $p_{klmn}$, where the indexes indicate the number of photons detected at the CHS ($k$), Alice's and Bob's stations ($l$ and $m$) and undetected ($n$), respectively.

Another source of errors is the indistinguishability of two photons. This is directly related to the Hong-Ou-Mandel (HOM) visibility $V_{\text{HOM}}$ \cite{eva2021bell}: the higher the visibility, the more indistinguishable the photons are. The precise nature of the indistinguishability, on the other hand, depends on the decoherence mechanisms of the single-photon source. Since in this work we consider implementations using single-photon sources based on quantum dots, we focus on fast (pure) dephasing with a rate $\gamma_d$, e.g. due to phonon interaction with the quantum dot, as this is the main source of noise for quantum dots on the relevant timescale \cite{petru}. The visibility and dephasing rate can then be shown to be related by \cite{eva2021bell}
\begin{equation}
V_{\text{HOM}} =  \frac{\gamma}{\gamma+2\gamma_d}\,,  
\label{eq:alpha_sq}
\end{equation}
where $\gamma$ is the spontaneous decay rate of the quantum-dot. Additionally, slow drifts can make Alice and Bob's emitters go in and out of resonance over time. We neglect this effect for simplicity and assume that all photons are equally indistinguishable. In addition, we assume that the limited visibility is only due to the fast dephasing rate $\gamma_d$. This assumption requires two remote quantum dots to have a similar visibility as that of a single quantum dot. This was demonstrated in Ref.~\cite{zhai2022}, but even if this is not the case, Ref.~\cite{eva2021bell} showed that, in practice, the performance of the protocol has a weaker dependence on the cross visibility between Alice and Bob quantum dots than their intrinsic one.

Finally, we consider multi-photon emission, as characterised by the second order correlation function for the pulse
\begin{equation}\label{Eq:g2_result}
    g^{(2)}=\frac{2P_2}{\left( P_1 + 2P_2 \right)^2}\,,
\end{equation}
for a multiphoton state described by the density matrix
\begin{equation}
    \hat{\rho}_s = \sum_k P_k \hat{\rho}_k\,,
    \label{eq:number_state}
\end{equation}
where $\hat{\rho}_k$ represents the $k$-th photon component state with probability $P_k$ and where $\sum_k P_k$ = 1. Note that the incoherent photon-number state assumed in \eqref{eq:number_state} is expected both when the quantum dot is excited above resonance, and with resonant excitation followed by active phase scrambling \cite{bozzio2022}.
We assume that $\hat{\rho}_s \simeq P_0 \hat{\rho}_0 + P_1 \hat{\rho}_1 + P_2 \hat{\rho}_2 + \mathcal{O}(P_3)$, since for quantum-dot single-photon sources the probability of generating more than one photon is low. In fact, we approximate all results to first order in $P_2$, such that we consider it to be very unlikely that two sources produce a two-photon state simultaneously ($P_2^2\simeq0$).  
Moreover, the second-generated photon is assumed to be completely distinguishable from the other four as typically done in the context of single-photon sources. We note, however, that this may not be the case; a full investigation of this is beyond the scope of the present article \cite{johannes,ollivier2021}. The presence of an extra photon in the set up thus induces an error that is effectively equivalent to having a detector dark count. 

\section{DIQKD protocol}\label{Sec:DIQKD}

In the remaining of this paper, the following assumptions are considered:
\begin{enumerate}
    \item Alice and Bob are both in secure laboratories and can thus prevent unwanted information from leaking outside of their respective locations. This means that their inputs and outputs remain secret until the public communication step.
    
    \item Alice and Bob can generate trusted local randomness, as for instance, when performing their measurement choice.
    
    \item Alice and Bob have an authenticated public channel.
\end{enumerate}

Under these assumptions, we consider the scenario described in the previous section, where Alice and Bob are in secure laboratories and want to share a secret key. Furthermore, Alice and Bob only record their measurement when they receive a \emph{yes} from the CHS. In our implementation, we consider a 2322 scenario where Alice has two measurement settings $x~\in \{0,1\}$ and Bob has three settings $y~\in \{0,1,2\}$ (specified by the classical inputs $\{x_d,y_d\}$), while both of them get two possible outcomes. In the following, we denote by $x'$ and $y'$ the key generating inputs, and we consider that an eavesdropper Eve has access to arbitrary quantum resources. The asymptotic rate $r$ of a DIQKD protocol
with one-way error correction is given by the Devetak-Winter bound \cite{devetak_distillation_2005}
\begin{equation}
r\geq H(A|X=x',E)-H(A|B,X=x',Y=y'),
\end{equation}
where $H(A\vert B)$ stands for the conditional von Neumann entropy of $A$ given $B$. 

From the statistics coming from the set-up described in the previous section, one can easily compute $H(A|B,X=x',Y=y')$. Thus, the main difficulty arises in bounding $H(A|X=x',E)$ knowing some statistical test $p(a,b|x,y)$. 
In Ref.~\cite{Acin07}, an analytical lower bound on $H(A|X=x',E)$ was derived by considering a particular statistical test which is given by the CHSH inequality \cite{Clauser69}. The robustness of the protocol with respect to losses has been improved using noisy preprocessing on Alice's side which can increase $H(A|X=x',E)$ more than $H(A|B,X=x',Y=y')$ (as described below) \cite{ho_noisy_2020}. Recently, the authors of Refs.~\cite{brown_device-independent_2021,brown_device-independent_2021_arxiv} derived a convergent series of lower bounds on $H(A|X=x',E)$ where the statistical test is given by the value of the probabilities $p(a,b|x,y)$ 
\begin{equation}\label{Eq:Brown:Entropy}
    \begin{aligned}
    H(A\vert X=x',E)
        \geq c_m
            &+ \sum_{i=1}^{m-1} \dfrac{w_i}{t_i \ln{2}}
            \\
            &\times\sum_{a} \inf_{Z_a\in B(E)}
                \expval{f(t_i, M_{a\vert x},Z_{a,i})},
    \end{aligned}
\end{equation}
where the function $f(t_i, M_{a\vert x},Z_{a,i})$ is defined as
\begin{equation}
    \begin{aligned}
    &f(t_i, M_{a\vert x},Z_{a,i})= 
        \\
        &\hspace{1cm}
         M_{a\vert x}
            \big(
                Z_{a,i}
                +Z^*_{a,i}
                +(1-t_i)Z_{a, i}^*Z_{a,i}
            \big)
            + t_i Z_{a,i}^*Z_{a,i}^*
    \end{aligned}
\end{equation}
where $w_i$ and $t_i$ are the $i$th Gauss-Radau quadratures and $c_m=\sum_{i=1}^{m-1} \frac{w_i}{t_i \ln{2}}$, such that increasing $m$ leads to better bounds (see Refs.~\cite{brown_device-independent_2021,brown_device-independent_2021_arxiv} for details), and $M_{a\vert x}$ and $N_{b\vert x}$ represent Alice's and Bob's POVM elements respectively, and $Z_{a,i}$ describes Eve's operators. In this expression, the following constraints need to be satisfied $\forall~i\in\{0,1\}, ~\forall~x\in\{0,1\}, ~\forall~y\in\{0,1,2\}$:
\begin{equation}
\begin{split}
    &\langle M_{a\vert x} N_{b\vert y}\rangle = p(a,b\vert x,y)\\
    &\langle M_{a\vert x}\rangle = p(a\vert x),
    \quad \langle N_{b\vert y} \rangle = p(b\vert y),
    \\
    &
        M_{a\vert x} > 0, \quad N_{b\vert y} >0,
        M_{a\vert x}, N_{b\vert y}~\text{and}~Z_{a,i}\in \text{B(H)}.
\end{split}
\end{equation}

The positivity of the POVM together with the condition $M_{a|x},~N_{b,y}\text{ and } Z_{a,i} \in \text{B(H)}$, where $\text{B(H)}$ is the ensemble of bounded operators, can be relaxed using the Navascues-Pironio-Acín hierarchy (NPA) \cite{navascues_bounding_2007,navascues_convergent_2008}. In the following, we refer to such relaxation by $\text{OPT}(m,A,B,x,y,t)$ with $t$ the transmissivity of the beam splitter located in the CHS. Note that, usually, Alice uses her two settings both to test CHSH and generate the key, while Bob uses his two first settings to test CHSH and the third to generate the key. Here, we slightly changed the scenario by allowing Bob to use his three settings to compute an additional joint probability which will sometimes be used as an additional statistical test. We aim to find the best experimental parameters for DIQKD using single photon sources. To this end, we perform the numerical optimization
\begin{equation}\label{eq:opt:rate}
    \begin{aligned}
    r& =\max_{x,y,t}
        \left(\text{OPT}(m,A,B,x,y,t)
            -H(A|B,X=x^{\prime},Y=y^{\prime})\right),
    \end{aligned}
\end{equation}
where the maximization runs over the possible experimental settings $x$ and $y$, which in our case are determined by the measurement angles of Alice ($\theta_{A,0}, \theta_{A,1}$) and over the three measurement angles of Bob ($\theta_{B,0}, \theta_{B,1}, \theta_{B,2}$), and over the transmissivity $t$ of the beam splitter at the CHS.

\section{Results}\label{Sec:Results}
In this section we present the results we obtained by solving the optimization problem presented in \eqref{eq:opt:rate}. We first compare it with other analytical techniques that have been developed in the literature \cite{ma_improved_2012,ho_noisy_2020,brown_device-independent_2021,brown_device-independent_2021_arxiv}, and then study the dependence of the key rate on the visibility $V_\text{HOM}$ and the autocorrelation function $g^{(2)}$. Finally, we consider the scenario where Alice and Bob only have  access to a limited number of rounds for generating the key. The details of the numerical implementation can be found in the Supplementary Material. 

\subsection{Comparison with analytical formulas}
Here, we compare the results of the optimization problem in \eqref{eq:opt:rate} with the analytical bounds presented in Refs.~\cite{ma_improved_2012,ho_noisy_2020}. In both references, the bound is obtained by considering a similar 2322 scenario where Alice performs two measurements, Bob performs three measurements and they both get two outcomes. Two of Bob's measurements are used to optimize the CHSH inequality, while the outcomes of the third are used to construct the key between both parties. By doing this, Ref.~\cite{ma_improved_2012} found the following bound on the key rate
\begin{equation}\label{Eq:CHSH:bound}
    r \geq 1
        - h\bigg(
                \dfrac{1+\sqrt{(S/2)^2-1}}{2}
            \bigg)
        - H(B_2\vert A_0),
\end{equation}
where $h(\cdot)$ represents the binary entropy and $S$ corresponds to the CHSH score
\begin{equation}
\begin{split}
       S = E(x_0,y_0) + E(x_1,y_0) + E(x_0,y_1) - E(x_1,y_1)\,, 
\end{split}
\end{equation}
and where $E(x_d,y_q) = \sum_{i,j} (-1)^{i\oplus j}  p(a_i,b_j|x_d,y_q)$.

In Ref.~\cite{ho_noisy_2020} the authors considered a different approach under which the data undergoes a preprocessing operation before being turned into the key. In particular, Alice performs a bit-flip operation on her outputs with probability $q$, which is equivalent to a change of her measurement operators from the set $\{M_0,M_1\}$ to $\{(1-q)M_0+qM_1,qM_0+(1-q)M_1\}$. Therefore, the bound of the key rate is given by
\begin{equation}\label{Eq:preprocessing:bound}
    \begin{aligned}
    r \geq 1
    &-h\bigg(
        \dfrac{1+\sqrt{(S/2)^2-1}}{2}
      \bigg)
     -H(B_2\vert A_0)
    \\
    &+h\bigg(
            \dfrac{1+\sqrt{1-q(q-1)(8-S^2)}}{2}
      \bigg),
    \end{aligned}
\end{equation}
where one can see that, depending on the CHSH score $S$, the introduction of the bit-flip operation $q$ may increase Eve's ignorance on Alice's output more than Bob's. This strategy thus allows enhancing the rate obtained in \eqref{Eq:CHSH:bound}. The same preprocessing technique can also be applied to our optimization problem in \eqref{eq:opt:rate}. The implementation of it, however, increases the number of variables we have to consider in the optimization.

\begin{figure}
    \centering
    \includegraphics[width = 1.\columnwidth]{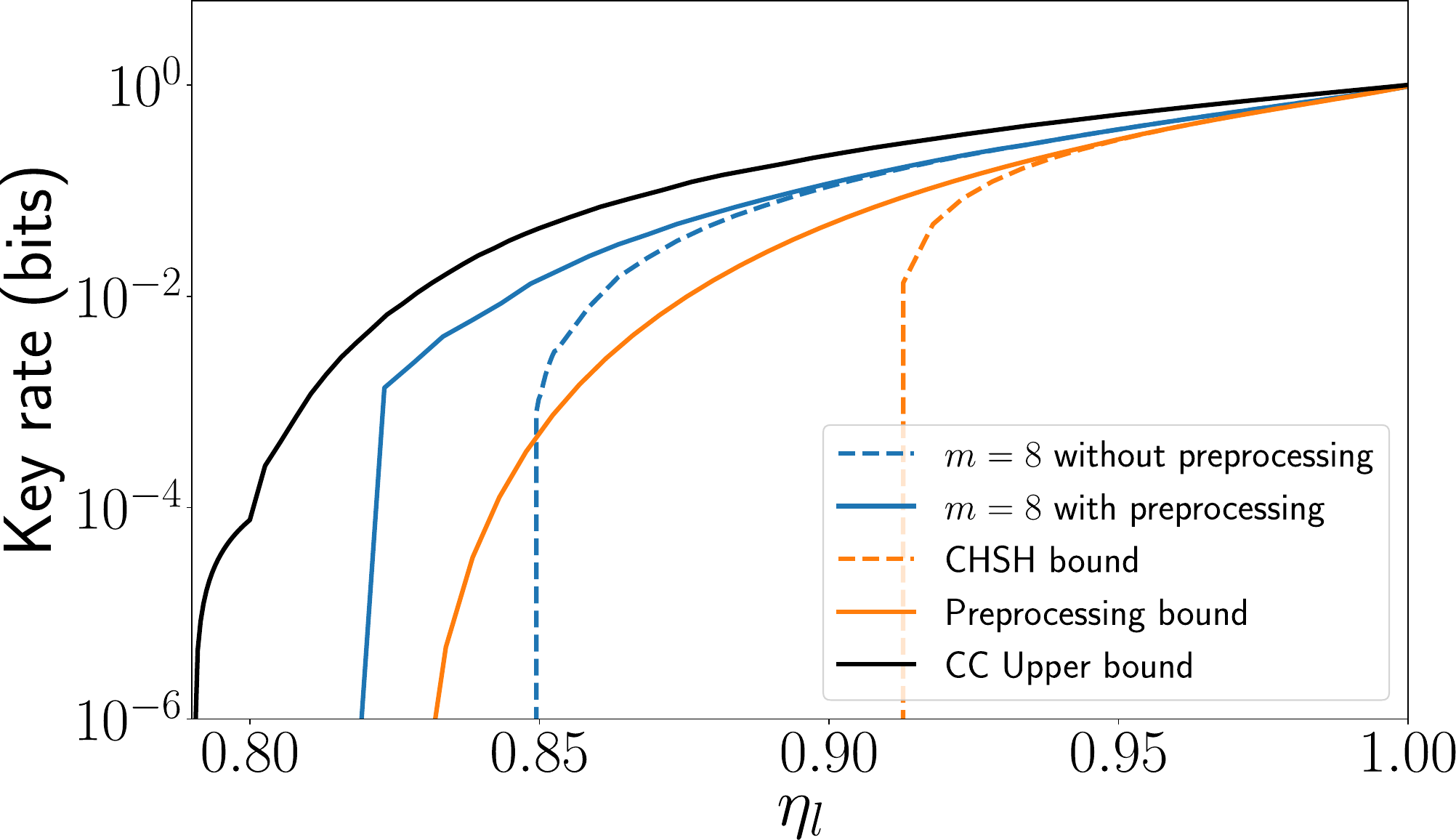}
    \caption{Dependence of the key rate on the detection efficiency $\eta_l$ for the case of $V_{\text{HOM}} = 1.0$ and $g^{(2)}=0$. Here, we have set the transmission efficiency to $T= 10^{-3}$. In blue, we show the key rate obtained by solving the optimization problem in \eqref{eq:opt:rate}, where the solid (dashed) curve shows the rate obtained by (not) implementing the preprocessing strategy. In orange, we show the results obtained by using the analytical formulas in Eqs.~\eqref{Eq:CHSH:bound} (dashed curve) and \eqref{Eq:preprocessing:bound} (solid curve). The use of the optimization technique in general leads to better bounds on the obtained key rate, compared to the corresponding analytic bounds. These results coincide with those obtained in Refs.~\cite{brown_device-independent_2021,brown_device-independent_2021_arxiv}. Finally, with the black solid curve we show an upper bound on the Devetak-Winter bound obtained by implementing the so-called Convex-Combination (CC) attack \cite{farkas_bell_2021,lukanowski_upper_2022} using one-way communication reconciliation techniques (the data of which was obtained upon request to the respective authors). }
    \label{fig:KeyRate:Vis1}
\end{figure}

With the aim of comparing the different techniques, we show in Fig.~\ref{fig:KeyRate:Vis1} the results for the ideal scenario $V_\text{HOM}=1.0$ and $g^{(2)} = 0.0$, i.e., where the photons generated by the quantum dots are completely indistinguishable and the generated state is pure. For now we consider the secret key rate per successful heralding event at the CHS. Hereupon, and unless otherwise stated, we therefore set the transmission efficiency to $T=10^{-3}$ to investigate the limit where it has a negligible influence on the local efficiency. This will allow us to  identify the minimal experimental requirements for obtaining a secret key  
\cite{eva2021bell}. We will later relax this assumption and investigate the influence of a higher transmission.

The orange curves correspond to the bounds obtained from \eqref{Eq:CHSH:bound} (orange dashed curve) and \eqref{Eq:preprocessing:bound} (orange solid curve), while blue curves show the solution found for the optimization problem presented in \eqref{eq:opt:rate} when setting $m=8$, with and without noisy preprocessing (blue solid and dashed curves, respectively). Note that in the following and unless otherwise stated, we always set $m=8$ in \eqref{eq:opt:rate} when evaluating the relative entropy between Alice and Eve. We see that, in both cases, the improved bound for the relative entropy between Alice and Eve shown in \eqref{Eq:Brown:Entropy} allows for enhancing the final value of the key rate and lowers the limits for when this quantity starts to be positive. In particular, for the situation without noisy preprocessing we find that the key rate starts to be positive for $\eta_l > 0.909$, according to the analytic bound in \eqref{Eq:CHSH:bound}, while the solution to \eqref{eq:opt:rate} leads to $\eta_l > 0.848$. As $\eta_l$ increases, the difference between both results becomes smaller, and they converge at $\eta_l = 1$ where $r \approx 1$ bit. On the other hand, the implementation of the noisy preprocessing strategy leads to an improvement with respect to the bound in \eqref{Eq:CHSH:bound}, and provides positive values of the key rate for $\eta_l > 0.832$. Introducing this strategy at the level of \eqref{eq:opt:rate} further reduces this limit to $\eta_l > 0.818$. Nevertheless, while the use of the noisy preprocessing truly improves the obtained bound on the efficiency, the differences in the key rate we get for $V_\text{HOM} = 1.0$ are almost negligible in comparison with the ones obtained without implementing it. For instance, at $\eta_l = 0.848$, when the rate for the latter is already zero, the rate obtained with the noisy preprocessing is $r\approx 10^{-3}$ bits. We expect this improvement to be negligible in real experimental implementations, where finite statistics have to be considered. Finally, with the black curve we show an upper bound to the Devetak-Winter bound obtained by implementing the so-called Convex-Combination attack \cite{farkas_bell_2021,lukanowski_upper_2022}. As seen in the figure, our results are close to the upper bound. 
This upper bound imposes a lower limit on the minimal value of $\eta_l$ (in this case $\eta_l = 0.791$) for which one could potentially find positive key rates using one-way communication reconciliation techniques. 

\subsection{Dependence of the key rate on the distinguishability of the generated photons}
As mentioned before, the HOM-visibility $V_\text{HOM}$ determines how distinguishable the photons are.
In state-of-the-art experimental implementations with quantum-dot single-photon sources, the values of the visibility that can be achieved are around 96\% \cite{Tomm:2021vq,Uppueabc8268} (note that this value contains a reduction due to $g^{(2)}$ so that the intrinsic visibility value is higher). This parameter affects the quality of the generated states, leading to a reduction of the CHSH inequality violation for a fixed value of the local efficiency \cite{eva2021bell}. Thus, we expect its role to be fundamental for DIQKD implementations.

\begin{figure}
    \centering
    \includegraphics[width = 1.\columnwidth]{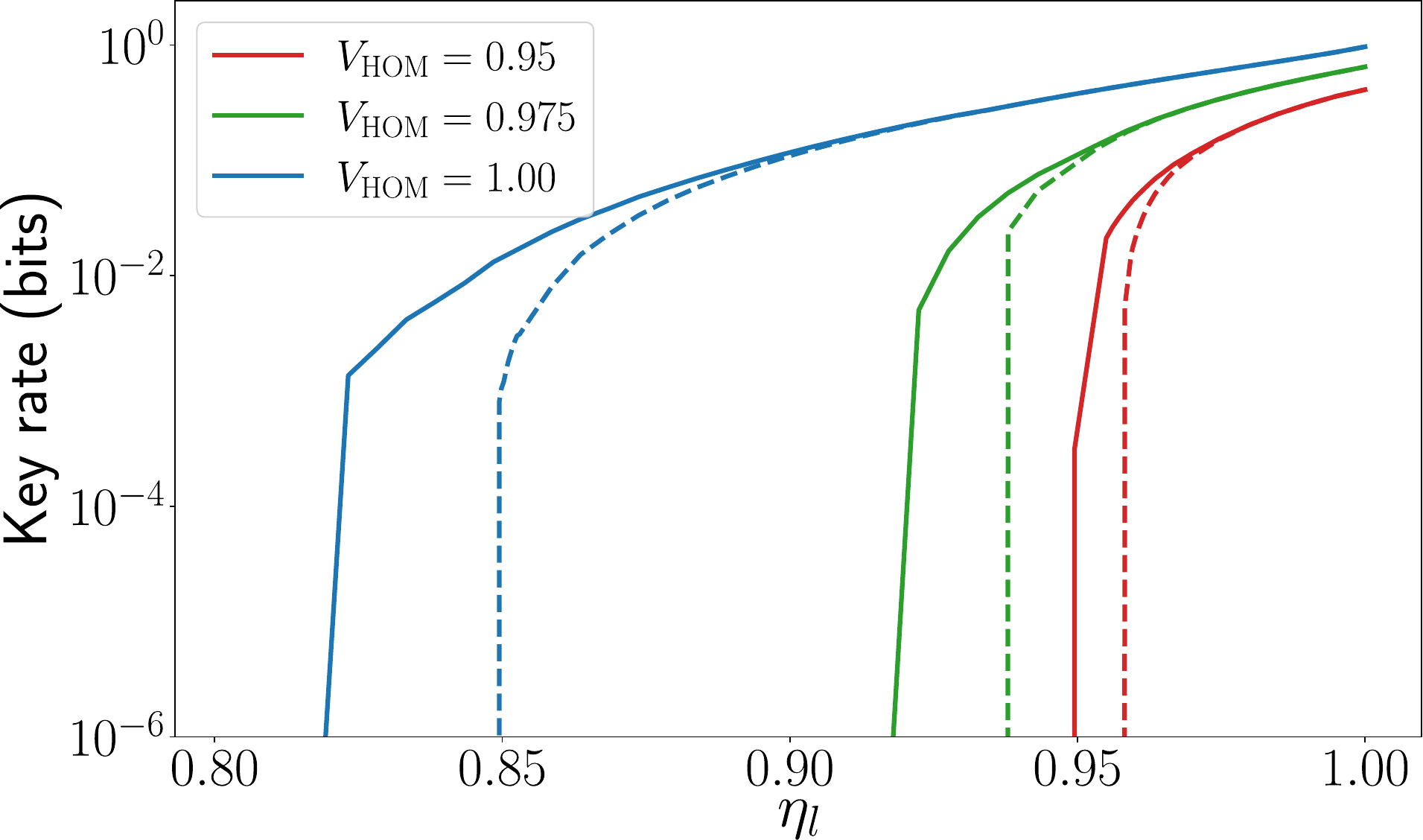}
    \caption{Dependence of the key rate on the visibility $V_{\text{HOM}}$. Here, we have set $g^{(2)} = 0.0$ and the transmission efficiency to $T=10^{-3}$. The solid curves show the key rate achieved when a noisy preprocessing strategy has been used for the generation of the key, while the dashed curves consider the case where the preprocessing has not been applied. In all curves, we have solved the optimization problem in \eqref{eq:opt:rate} setting $m=8$ and going up to second level in the NPA hierarchy (see Supplementary Material for further details).}
    \label{Fig:Key:and:Vis}
\end{figure}

In Fig.~\ref{Fig:Key:and:Vis}, we show the dependence of the key rate on the local efficiency for three different values of the visibility, namely $V_{\text{HOM}} = 0.95$ (red curves), $V_{\text{HOM}} = 0.975$ (green curves) and $V_{\text{HOM}} = 1.00$ (blue curves). In all cases we set $g^{(2)} = 0.00$. For a given value of the visibility, we show two curves: the dashed curve without the noisy preprocessing strategy, and the solid curve for which this strategy has been applied. As we observe, the introduction of the preprocessing strategy in general allows enhancing the key rate for small values of the local efficiency $\eta_l$, lowering the value of $\eta_l$ for which the key starts to be positive. In particular, for $V_{\text{HOM}} = 0.975$ we find that the key rate starts to be positive for $\eta_l > 0.918$ and $\eta_l > 0.938$ with and without the noisy preprocessing step respectively, while for $V_{\text{HOM}} = 0.95$ we get $\eta_l > 0.949$ and $\eta_l > 0.958$. On the other hand, as $\eta \to 1$, the solid and dashed curves converge to the same value, and lead to $r = 0.648$ for $V_{\text{HOM}} = 0.975 $, and $r = 0.409$ for $V_{\text{HOM}} = 0.95$. 

\subsection{Dependence of the key rate on the purity 
of the generated state}
The probability of generating a second photon from the source can be quantified in terms of the second order autocorrelation function, as shown in \eqref{Eq:g2_result}.
In experimental implementations, the degree of purity that can be attained is typically very high $g^{(2)} \leq 0.05$ \cite{Tomm:2021vq,Uppueabc8268}. Nevertheless, and as it happens with the visibility, this quantity is also fundamental for establishing nonlocal correlations between Alice and Bob \cite{eva2021bell}, and therefore for generating a secure key between both parties.

\begin{figure}
    \centering
    \includegraphics[width = 1.\columnwidth]{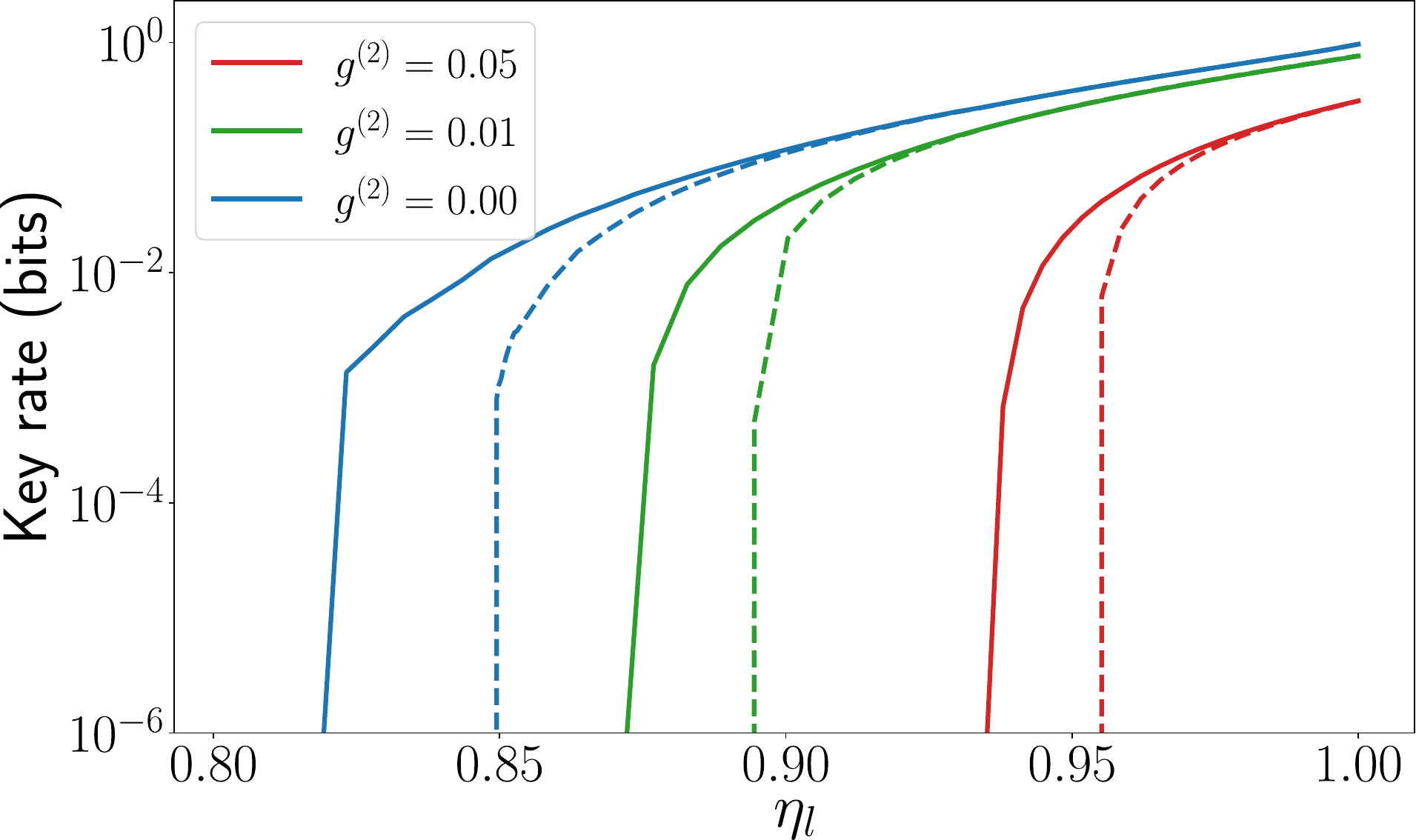}
    \caption{Dependence of the key rate $r$ on the  autocorrelation function $g^{(2)}$ in log scale. Here, we have set $V_{\text{HOM}} = 1.0$ and the transmission efficiency to $T=10^{-3}$. The solid curves show the key rate achieved when a noisy preprocessing strategy has been used for the generation of the key, while the dashed curves consider the case where the preprocessing has not been applied. In all these curves, we have solved the optimization problem in \eqref{eq:opt:rate} setting $m=8$ and going up to second level in the NPA hierarchy (see Supplementary Material for more information).}
    \label{Fig:Key:and:g2}
\end{figure}

In Fig.~\ref{Fig:Key:and:g2}, we show the dependence of the key rate on the local efficiency for three different values of the autocorrelation function, $g^{(2)} = 0.05$ (red curves), $g^{(2)} = 0.01$ (green curves) and $g^{(2)} = 0.00$ (blue curves). Here, we have set $V_{\text{HOM}} = 1.00$. Similar to the visibility analysis, we show the situation where the noisy preprocessing strategy is (not) applied with solid (dashed) curves. We observe that, as it happens with the HOM-visibility, small modifications in the $g^{(2)}$ autocorrelation function have important consequences on the requirements for the detection efficiencies needed for obtaining positive values of the key rate. The noisy preprocessing strategy is, however, again useful for relaxing these requirements. In particular, we observe that for $g^{(2)} = 0.01$ we get positive key rates for $\eta_l > 0.872$ and $\eta_l > 0.894$ with and without the noisy preprocessing, respectively. On the other hand, for $g^{(2)} = 0.05$ the key rate starts to be positive for $\eta_l > 0.935$ and $\eta_l > 0.955$. Furthermore, in the limit of large efficiency ($\eta_l \to 1$), the solid and dashed lines converge to the same value, which is $r = 0.766$ for $g^{(2)} = 0.01$ and $r = 0.311$ for $g^{(2)} = 0.05$.

From the analysis we have done in this section, we observe that in the asymptotic limit of infinite rounds of the protocol, the main limiting factor in current experimental implementations lies in the local detection efficiency $\eta_l$. The limiting value of $\eta_l$ for which the key rate becomes positive can be improved by means of noisy preprocessing techniques. For large values of $\eta_l$ the difference with the \emph{standard} protocol is, however, limited. In the range of values for the visibility and second order autocorrelation function that is currently accessible experimentally for quantum dot single-photon sources ($V_{\text{HOM}} \sim 0.96$, $g^{(2)} \leq 0.05$ \cite{Tomm:2021vq,Uppueabc8268}), the key established between Alice and Bob starts to be secure for $\eta_l \gtrsim 0.9$. In the following, we consider a more realistic scenario where Alice and Bob have access to a limited amount of rounds $n$ for the protocol, which adds an extra constraint to realistic DIQKD implementations.

\subsection{Finite size analysis}
Alice and Bob now only have access to a finite number of rounds $n$, which reduces the length $l$ of secure key they can generate. On top of that, both parties are far away one from the other, and thus the photons that are sent to the CHS may be absorbed by the medium. In this section, we take these two experimental limitations into account. This allows us to provide a bound on the time and distance for which a secret key can be shared between Alice and Bob in photonic DIQKD experiments with quantum dots. 
\begin{figure}
    \centering
    \includegraphics[width =\linewidth]{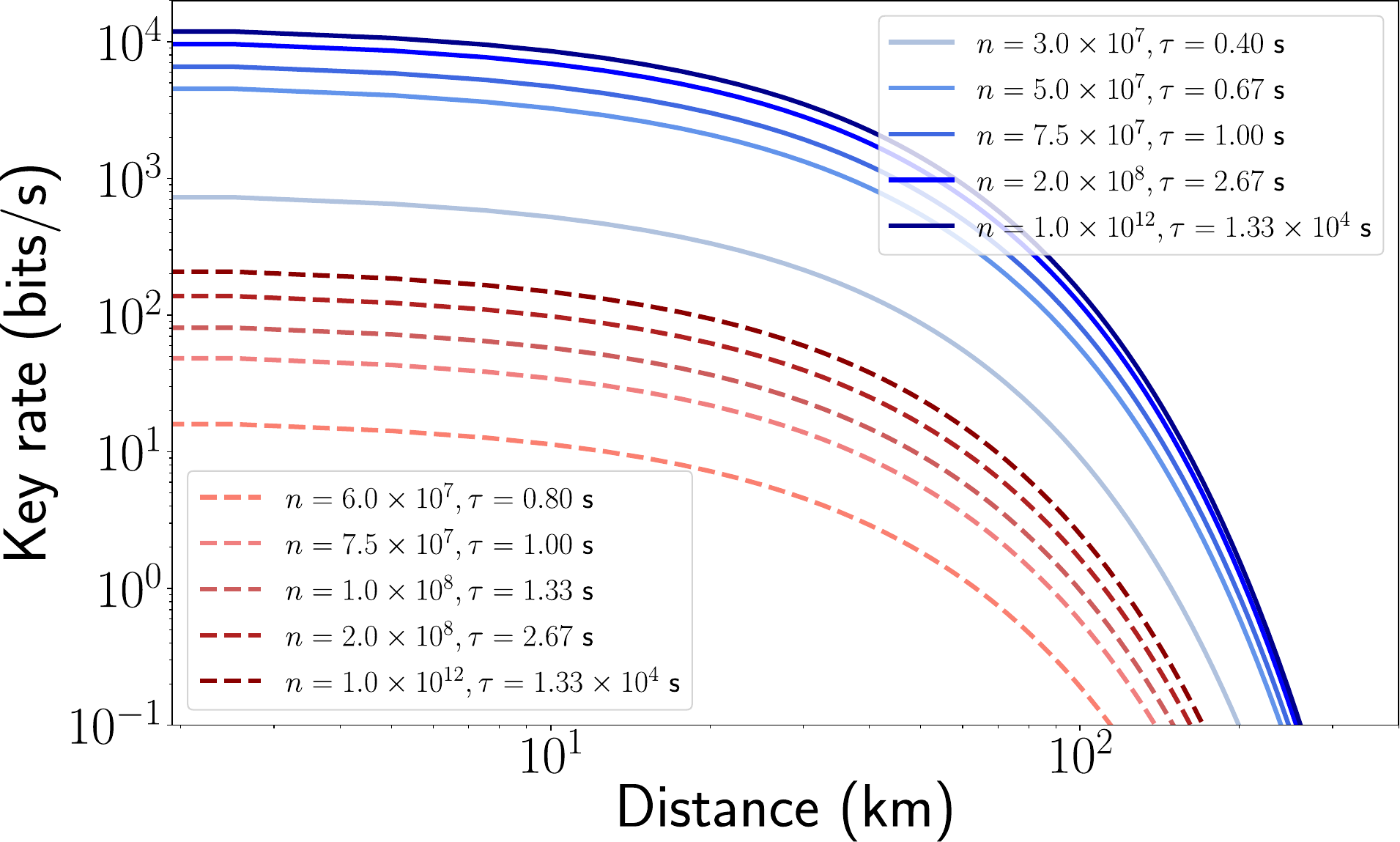}
    \caption{Key rate in bits per second as a function of the distance. We consider a single-photon source that generates photons at a rate of $\nu = 75$ MHz. The blue solid curves show the ideal case where $\eta_l = 1.0$, $V_{\text{HOM}} = 1.0$ and $g^{(2)}=0.00$. The red dashed curves show a more realistic case where $\eta_l = 0.957$, $V_{\text{HOM}} = 0.975$ and $g^{(2)}=0.01$. Each of the curves shown in the plot are obtained for different number of rounds $n$ (see legend), such that the time required to perform this number of rounds is obtained as $\tau = n/\nu$. Here we have optimized the transmissivity of the beam splitter $T$ for each set of experimental parameters. The optimal values are $T=6\%$ for the blue solid curves, and $T= 1\%$ for the red dashed curves.}
    \label{Fig:Key:Rounds}
\end{figure}

A recent breakthrough regarding the computation of key rates for finite statistics is the entropy accumulation theorem (EAT) \cite{dupuis20}. This theorem allows us to provide a security proof of the protocol against general attacks, provided that newly generated side information at each round follows the Markov condition. The different steps of the protocol can be slightly changed such that this hypothesis is satisfied \cite{tan_improved_2020}. The EAT bounds the amount of randomness generated by two parties ($\text{Rand}_{\text{Alice-Bob}}$) in $n$ rounds against an arbitrary powerful quantum adversary Eve given a statistical test, here the CHSH score $S$, according to 
\begin{equation}
    \text{Rand}_{\text{Alice-Bob}}\geq n h(S) -\sqrt{n}k
\end{equation}
where $h(S)$ is the worst-case von Neumann entropy of an
individual round of the protocol with expected score $S$, and $k$ is a correction factor which comes from the finite statistics. The length of the key is then given by the difference between the amount of entropy certified by the EAT, and the amount of information shared in the classical message during privacy amplification and error correction. We follow the approach of Ref.~\cite{tan_improved_2020}, for bounding the single round von Neumann entropy \cite{Liu21,nadlinger_experimental_2022}. Therefore, Theorem 1 in Ref.~\cite{tan_improved_2020} allows us to obtain an upper bound on the length $l$ of the secure key that we are able to generate with the presented protocol. The length $l$ is linked to the key rate by the number of protocol rounds $n$ that Alice and Bob undergo, i.e. $r = l/n$, such that in the asymptotic limit $n\to\infty$ we recover \eqref{Eq:CHSH:bound}. For more details about the extensive content of Theorem 1 and the parameters it encompasses, we refer the reader to page 14 of Ref.~\cite{tan_improved_2020}. 

In realistic experimental implementations, the number of rounds is usually determined by the rate at which the single-photon source generates photons, and the amount of time Alice and Bob spend running the protocol to generate the key. In typical quantum dot implementations, the rate is typically around $\nu = 75$ MHz \cite{Uppueabc8268,Tomm:2021vq,wang_towards_2019}. Furthermore, in order to consider one out of the $n$ protocol rounds to be successful, Alice and Bob need to get a positive response from the CHS. Therefore, the distance between both parties is a key aspect to take into account. Large distances between Alice and Bob lead to a decreasing probability of heralding, as it is more likely for the heralding photon to get lost on its route to the CHS. In general, the transmission efficiency can be written as $\eta_t = \eta_{\text{conv}}\exp{-\tfrac{L}{2L_0}}$,
where $L$ is the distance between Alice and Bob, $L_0$ is the attenuation length of the media through which the light travels (the factor of two arises because the distance to the CHS is only half the distance between Alice and Bob), and $\eta_{\text{conv}}$ is the efficiency conversion to telecommunication wavelength. The wavelength conversion is needed as InGaAs quantum dots typically emit at 947 nm, while telecommunication optical fibers operate at either 1300 nm or 1550 nm. We set the efficiency of this process to $\eta_{\text{conv}}=0.5$, as shown experimentally \cite{dalio2022,zahidy2023}. For optical fibers at the telecommunication wavelength a typical value is $L_0 = 22$ km. 

The transmission efficiency $\eta_t$ affects the probability of heralding $P_h$, which we get by summing all the probability terms that correspond to acceptance from the CHS (see Table S1 in Supplementary Material). These terms already account for all possibilities of emission of a second-generated photon when $g^{(2)}\neq0$. We note that this probability is as well a function of the beam splitter transmittance $T$ (see Fig.~\ref{Fig:CHS:setup}), such that small values of $T$ lead to low values of the heralding probability but to higher asymptotic key rates per successful heralding, as we reduce the chance of sending two photons to the CHS. On the other hand, large values of $T$ increase the heralding probability at the cost of reducing the security of the key. Thus, in order to establish sufficiently long keys between Alice and Bob, the probability of heralding, and therefore the transmissivity $T$, plays a fundamental role.

In Fig.~\ref{Fig:Key:Rounds}, we show the dependence of the key rate (in bits per second) as a function of the distance for different experimental parameters when using a quantum dot single-photon source that generates photons with a rate of $\nu = 75$ MHz. We have followed Ref.~\cite{tan_improved_2020} in order to perform the finite size analysis, and we have imposed our protocol to be $10^{-6}$-sound and $10^{-3}$-complete, with the soundness and completeness parameters respectively representing the probability of generating an insecure key and the probability of aborting the protocol (see Ref.~\cite{tan_improved_2020} for definition). In particular, with the blue solid curves we consider an ideal scenario where $\eta_l = 1.0$, $V_{\text{HOM}} = 1.00$ and $g^{(2)} = 0.00$. 
On the other hand, the red dashed curves present a more realistic scenario with $V_{\text{HOM}} = 0.975$ and $g^{(2)} = 0.01$, achievable by current technologies \cite{Tomm:2021vq,Uppueabc8268}, and a local efficiency of $\eta_l = 0.957$ estimated from the results from Figs. \ref{Fig:Key:and:Vis} and \ref{Fig:Key:and:g2} in order to show a positive key rate.
Each of the curves shown in the figure correspond to a different number of protocol rounds $n$ used by Alice and Bob for generating the key (see legends). Here, $\tau = n/\nu$ corresponds to the time it takes a single-photon source that generates photons with rate $\nu = 75$ MHz to perform $n$ rounds of the protocol. For both kinds of curves (blue solid and red dashed), we observe that the larger $n$ is, the higher the key rate. Furthermore, in both cases we find that for $n\leq 10^{7}$ it is not possible to extract a key, according to the soundness and completeness requirements that we use. An increase in the number of rounds leads to larger values of key rate, up to a maximum value that corresponds to the asymptotic regime which we represent with the cases where $n=10^{12}$ (shown with the darkest solid and dashed curves). Running this number of rounds takes on the order of $\tau\sim10^4$ seconds when using a source that generates photons with a rate of $\nu=75$ MHz. However, we observe that with  $\tau\sim 1$ s, Alice and Bob could obtain key rates comparable to those found in the asymptotic regime. 

Unlike the plots shown in the previous section, in Fig.~\ref{Fig:Key:Rounds} we have considered an extra optimization over the transmissivity $T$ of the beam splitter in each of Alice's and Bob's stations, assuming them to be equal. This way, we can increase the probability of heralding $P_h$ up to the order of 10$^{-2}$, allowing us to obtain secure keys at longer distances after a limited time. In particular, we observe that in the worst of the considered scenarios, i.e. the red dashed curves, the key rate reduces to 0.1 bits/s at a range of distances $L \in [113,170]$ km (the limits obtained for $n = 6.0\times10^7$ and $n=10^{12}$ rounds respectively). On the other hand, for the ideal case we find the same key rate for distances $L\in[200,261]$ km (the limits obtained for $n=3.0\times 10^7$ and $n=10^{12}$ rounds respectively). Therefore, the increase of the heralding probability allows establishing secure rates for longer distances, at the cost of reducing the length of  key per  second. An alternative approach would be to increase the repetition rate $\nu$ of the single-photon source, although it might require more effort from the experimental side.

\section{Conclusion}
We have used new theoretical tools to evaluate the experimental requirements in order to perform DIQKD using single-photon sources in a heralded scheme. We focused on the parameters which are routinely used to characterize single-photon sources, such as the time-zero second order correlation function $g^{(2)}$, the HOM visibility $V_{\text{HOM}}$ and the local detection efficiency $\eta_l$ of the source. We have computed the asymptotic key rate using the novel method of Refs.~\cite{brown_device-independent_2021,brown_device-independent_2021_arxiv} for bounding the relative entropy between Alice and Eve, and we showed that this improves the analytical bounds obtained in Refs. \cite{ma_improved_2012,ho_noisy_2020}. Finally, we considered a situation where the number of rounds for obtaining a secure key rate is limited, i.e., a finite statistics analysis. We studied the time and distance requirements needed to establish a secure rate between Alice and Bob with realistic parameters obtainable with quantum dot single-photon sources. However, although we have focused on single-photon sources using quantum dots, the methods we have employed can be trivially extended to any other single-photon source.

Our analysis improves the experimental requirements for performing DIQKD with single-photon sources, and we reach numbers which are within reach for near-future experimental implementations. We believe that our work represents a strong motivation for the improvement of existing single-photon sources, so that they can be used to perform DIQKD. A way to implement this DIQKD protocol with single-photon sources that do not satisfy yet the exposed experimental requirements, could be to relax the assumption that Alice and Bob do not have knowledge on their device, or to constrain Eve to perform weaker attacks. Alternatively, the application of two-way information reconciliation techniques (instead of the one-way scenario presented in this work) may be explored. These adjustments could result in more lenient experimental requirements. In the latter case, such relaxed requirements have already been demonstrated for upper bounds on the key rate, as shown in Refs.~\cite{lukanowski_upper_2022} and \cite{zhang_quantum_2023}. Finally, the extension of the new bound developed by Ref.~\cite{brown_device-independent_2021} to a finite size key rate analysis would be the natural outlook to further extend the present analysis. This study could include also the influence of the transmittance T on the required key rate per successful event as yet another optimisation parameter. This could lead to finite-size key rates larger than the analytical CHSH results presented here.\\

\begin{backmatter}
\bmsection{Funding} Hy-Q acknowledges the support of Danmarks Grundforskningsfond (DNRF 139, Hy-Q Center for Hybrid Quantum Networks). ICFO group acknowledges support from the Government of Spain (FUNQIP, NextGenerationEU (PRTR-C17.I1) and Severo Ochoa CEX2019-000910-S), Fundació Cellex, Fundació Mir-Puig, Generalitat de Catalunya (CERCA program), EU project QSNP, the ERC AdG CERQUTE and the AXA Chair in Quantum Information Science. J.R-D. acknowledges support from the Secretaria d'Universitats i Recerca del Departament d'Empresa i Coneixement de la Generalitat de Catalunya, as well as the European Social Fund (L'FSE inverteix en el teu futur)--FEDER.

\bmsection{Acknowledgments} We gratefully thank K.~Łukanowski and J.~Kołodyński for providing us the data for the upper bound to the Devetak-Winter bound using the CC attack using one-way communication reconciliation techniques (black curve in Fig.~\ref{fig:KeyRate:Vis1}).

\bmsection{Disclosures} The authors declare no conflicts of interest.

\bmsection{Data availability} Data underlying the results presented in this paper are available in Ref.~\cite{ZenodoLink}.

\bmsection{Supplemental document}
See Supplementary Material for supporting content. 

\end{backmatter}


\bibliography{biblio}

\begin{thebibliography}{10}
\newcommand{\enquote}[1]{``#1''}

\bibitem{BB84}
C.~H. Bennett and G.~Brassard, \enquote{Quantum cryptography: {Public} key distribution and coin tossing,} {\protect\JournalTitle{Theoretical Computer Science}} \textbf{560}, 7--11 (2014). ArXiv:2003.06557 [quant-ph].

\bibitem{Ekert91}
A.~K. Ekert, \enquote{Quantum cryptography based on bell's theorem,} {\protect\JournalTitle{Phys. Rev. Lett.}} \textbf{67}, 661--663 (1991).

\bibitem{Scarani09}
V.~Scarani, H.~Bechmann-Pasquinucci, N.~J. Cerf, \emph{et~al.}, \enquote{The security of practical quantum key distribution,} {\protect\JournalTitle{Rev. Mod. Phys.}} \textbf{81}, 1301--1350 (2009).

\bibitem{Acin07}
A.~Ac\'{\i}n, N.~Brunner, N.~Gisin, \emph{et~al.}, \enquote{Device-independent security of quantum cryptography against collective attacks,} {\protect\JournalTitle{Phys. Rev. Lett.}} \textbf{98}, 230501 (2007).

\bibitem{Lydersen10}
L.~Lydersen, C.~Wiechers, C.~Wittmann, \emph{et~al.}, \enquote{Hacking commercial quantum cryptography systems by tailored bright illumination,} {\protect\JournalTitle{Nature Photonics}} \textbf{4}, 686--689 (2010).

\bibitem{Friedman19}
R.~Arnon-Friedman, R.~Renner, and T.~Vidick, \enquote{Simple and tight device-independent security proofs,} {\protect\JournalTitle{SIAM Journal on Computing}} \textbf{48}, 181--225 (2019).

\bibitem{Clauser69}
J.~F. Clauser, M.~A. Horne, A.~Shimony, and R.~A. Holt, \enquote{Proposed experiment to test local hidden-variable theories,} {\protect\JournalTitle{Phys. Rev. Lett.}} \textbf{23}, 880--884 (1969).

\bibitem{mayers_self_2004}
D.~Mayers and A.~Yao, \enquote{Self testing quantum apparatus,}  (2004). ArXiv:quant-ph/0307205.

\bibitem{Kaniewski16}
J.~Kaniewski, \enquote{Analytic and nearly optimal self-testing bounds for the clauser-horne-shimony-holt and mermin inequalities,} {\protect\JournalTitle{Phys. Rev. Lett.}} \textbf{117}, 070402 (2016).

\bibitem{nadlinger_experimental_2022}
D.~P. Nadlinger, P.~Drmota, B.~C. Nichol, \emph{et~al.}, \enquote{Experimental quantum key distribution certified by {Bell}'s theorem,} {\protect\JournalTitle{Nature}} \textbf{607}, 682--686 (2022).

\bibitem{Lim21}
W.~Zhang, T.~van Leent, K.~Redeker, \emph{et~al.}, \enquote{A device-independent quantum key distribution system for distant users,} {\protect\JournalTitle{Nature}} \textbf{607}, 687--691 (2022). Number: 7920 Publisher: Nature Publishing Group.

\bibitem{Sangouard10}
N.~Gisin, S.~Pironio, and N.~Sangouard, \enquote{Proposal for implementing device-independent quantum key distribution based on a heralded qubit amplifier,} {\protect\JournalTitle{Phys. Rev. Lett.}} \textbf{105}, 070501 (2010).

\bibitem{main_article}
J.~Ko{\l{}}ody{\'{n}}ski, A.~M{\'{a}}ttar, P.~Skrzypczyk, \emph{et~al.}, \enquote{Device-independent quantum key distribution with single-photon sources,} {\protect\JournalTitle{{Quantum}}} \textbf{4}, 260 (2020).

\bibitem{Tomm:2021vq}
N.~Tomm, A.~Javadi, N.~O. Antoniadis, \emph{et~al.}, \enquote{A bright and fast source of coherent single photons,} {\protect\JournalTitle{Nature Nanotechnology}} \textbf{16}, 399--403 (2021).

\bibitem{Uppueabc8268}
R.~Uppu, F.~T. Pedersen, Y.~Wang, \emph{et~al.}, \enquote{Scalable integrated single-photon source,} {\protect\JournalTitle{Science Advances}} \textbf{6} (2020).

\bibitem{Sekatski21}
P.~Sekatski, J.-D. Bancal, X.~Valcarce, \emph{et~al.}, \enquote{Device-independent quantum key distribution from generalized {CHSH} inequalities,} {\protect\JournalTitle{{Quantum}}} \textbf{5}, 444 (2021).

\bibitem{Woodhead21}
E.~Woodhead, A.~Ac{\'{i}}n, and S.~Pironio, \enquote{Device-independent quantum key distribution with asymmetric {CHSH} inequalities,} {\protect\JournalTitle{{Quantum}}} \textbf{5}, 443 (2021).

\bibitem{brown_device-independent_2021}
P.~Brown, H.~Fawzi, and O.~Fawzi, \enquote{Computing conditional entropies for quantum correlations,} {\protect\JournalTitle{Nature Communications}} \textbf{12}, 575 (2021).

\bibitem{brown_device-independent_2021_arxiv}
P.~Brown, H.~Fawzi, and O.~Fawzi, \enquote{Device-independent lower bounds on the conditional von {Neumann} entropy,} Tech. Rep. arXiv:2106.13692, arXiv (2021). ArXiv:2106.13692 [quant-ph].

\bibitem{araujo2022}
M.~Araújo, M.~Huber, M.~Navascués, \emph{et~al.}, \enquote{Quantum key distribution rates from semidefinite programming,}  (2022). ArXiv:2211.05725 [quant-ph].

\bibitem{farkas_bell_2021}
M.~Farkas, M.~Balanzó-Juandó, K.~Łukanowski, \emph{et~al.}, \enquote{Bell {Nonlocality} {Is} {Not} {Sufficient} for the {Security} of {Standard} {Device}-{Independent} {Quantum} {Key} {Distribution} {Protocols},} {\protect\JournalTitle{Physical Review Letters}} \textbf{127}, 050503 (2021).

\bibitem{tan_improved_2020}
E.~Y.-Z. Tan, P.~Sekatski, J.-D. Bancal, \emph{et~al.}, \enquote{Improved {DIQKD} protocols with finite-size analysis,} {\protect\JournalTitle{Quantum}} \textbf{6}, 880 (2022).

\bibitem{eva2021bell}
E.~M. Gonz\'alez-Ruiz, S.~K. Das, P.~Lodahl, and A.~S. S\o{}rensen, \enquote{Violation of bell's inequality with quantum-dot single-photon sources,} {\protect\JournalTitle{Phys. Rev. A}} \textbf{106}, 012222 (2022).

\bibitem{BSM}
P.~G. Kwiat and H.~Weinfurter, \enquote{Embedded bell-state analysis,} {\protect\JournalTitle{Phys. Rev. A}} \textbf{58}, R2623--R2626 (1998).

\bibitem{eberhard}
P.~H. Eberhard, \enquote{Background level and counter efficiencies required for a loophole-free einstein-podolsky-rosen experiment,} {\protect\JournalTitle{Phys. Rev. A}} \textbf{47}, R747--R750 (1993).

\bibitem{eta8284}
S.~Massar, S.~Pironio, J.~Roland, and B.~Gisin, \enquote{Bell inequalities resistant to detector inefficiency,} {\protect\JournalTitle{Phys. Rev. A}} \textbf{66}, 052112 (2002).

\bibitem{petru}
E.~A. Muljarov and R.~Zimmermann, \enquote{Dephasing in quantum dots: Quadratic coupling to acoustic phonons,} {\protect\JournalTitle{Phys. Rev. Lett.}} \textbf{93}, 237401 (2004).

\bibitem{zhai2022}
L.~Zhai, G.~N. Nguyen, C.~Spinnler, \emph{et~al.}, \enquote{Quantum interference of identical photons from remote {GaAs} quantum dots,} {\protect\JournalTitle{Nature Nanotechnology}} \textbf{17}, 829--833 (2022). Number: 8 Publisher: Nature Publishing Group.

\bibitem{bozzio2022}
M.~Bozzio, M.~Vyvlecka, M.~Cosacchi, \emph{et~al.}, \enquote{Enhancing quantum cryptography with quantum dot single-photon sources,} {\protect\JournalTitle{npj Quantum Information}} \textbf{8}, 1--8 (2022). Number: 1 Publisher: Nature Publishing Group.

\bibitem{johannes}
J.~Bjerlin, E.~M. González-Ruiz, and A.~S.~S{\o}rensen (2023). {In preparation}.

\bibitem{ollivier2021}
H.~Ollivier, S.~Thomas, S.~Wein, \emph{et~al.}, \enquote{Hong-ou-mandel interference with imperfect single photon sources,} {\protect\JournalTitle{Physical Review Letters}} \textbf{126}, 063602 (2021). Publisher: APS.

\bibitem{devetak_distillation_2005}
I.~Devetak and A.~Winter, \enquote{Distillation of secret key and entanglement from quantum states,} {\protect\JournalTitle{Proceedings of the Royal Society A: Mathematical, Physical and Engineering Sciences}} \textbf{461}, 207--235 (2005).

\bibitem{ho_noisy_2020}
M.~Ho, P.~Sekatski, E.-Z. Tan, \emph{et~al.}, \enquote{Noisy {Preprocessing} {Facilitates} a {Photonic} {Realization} of {Device}-{Independent} {Quantum} {Key} {Distribution},} {\protect\JournalTitle{Physical Review Letters}} \textbf{124}, 230502 (2020).

\bibitem{navascues_bounding_2007}
M.~Navascués, S.~Pironio, and A.~Acín, \enquote{Bounding the {Set} of {Quantum} {Correlations},} {\protect\JournalTitle{Physical Review Letters}} \textbf{98}, 010401 (2007).

\bibitem{navascues_convergent_2008}
M.~Navascués, S.~Pironio, and A.~Acín, \enquote{A convergent hierarchy of semidefinite programs characterizing the set of quantum correlations,} {\protect\JournalTitle{New Journal of Physics}} \textbf{10}, 073013 (2008).

\bibitem{ma_improved_2012}
X.~Ma and N.~Lütkenhaus, \enquote{Improved data post-processing in quantum key distribution and application to loss thresholds in device independent {QKD},} {\protect\JournalTitle{Quantum Information \& Computation}} \textbf{12}, 203--214 (2012).

\bibitem{lukanowski_upper_2022}
K.~Łukanowski, M.~Balanzó-Juandó, M.~Farkas, \emph{et~al.}, \enquote{Upper bounds on key rates in device-independent quantum key distribution based on convex-combination attacks,}  (2022). ArXiv:2206.06245 [quant-ph].

\bibitem{dupuis20}
F.~Dupuis, O.~Fawzi, and R.~Renner, \enquote{Entropy accumulation,} {\protect\JournalTitle{Communications in Mathematical Physics}} \textbf{379}, 867--913 (2020).

\bibitem{Liu21}
W.-Z. Liu, M.-H. Li, S.~Ragy, \emph{et~al.}, \enquote{Device-independent randomness expansion against quantum side information,} {\protect\JournalTitle{Nature Physics}} \textbf{17}, 448--451 (2021).

\bibitem{wang_towards_2019}
H.~Wang, Y.-M. He, T.-H. Chung, \emph{et~al.}, \enquote{Towards optimal single-photon sources from polarized microcavities,} {\protect\JournalTitle{Nature Photonics}} \textbf{13}, 770--775 (2019).

\bibitem{dalio2022}
B.~Da~Lio, C.~Faurby, X.~Zhou, \emph{et~al.}, \enquote{A {Pure} and {Indistinguishable} {Single}-{Photon} {Source} at {Telecommunication} {Wavelength},} {\protect\JournalTitle{Advanced Quantum Technologies}} \textbf{5}, 2200006 (2022). \_eprint: https://onlinelibrary.wiley.com/doi/pdf/10.1002/qute.202200006.

\bibitem{zahidy2023}
M.~Zahidy, M.~T. Mikkelsen, R.~Müller, \emph{et~al.}, \enquote{Quantum {Key} {Distribution} using {Deterministic} {Single}-{Photon} {Sources} over a {Field}-{Installed} {Fibre} {Link},}  (2023). ArXiv:2301.09399 [quant-ph].

\bibitem{zhang_quantum_2023}
X.~Zhang, P.~Zeng, T.~Ye, \emph{et~al.}, \enquote{Quantum {Complementarity} {Approach} to {Device}-{Independent} {Security},} {\protect\JournalTitle{Physical Review Letters}} \textbf{131}, 140801 (2023).

\bibitem{ZenodoLink}
E.~M. González-Ruiz, J.~Rivera-Dean, M.~F.~B. Cenni, \emph{et~al.}, \enquote{Device independent quantum key distribution with realistic single-photon source implementations,} \url{https://zenodo.org/record/7038040#.Y4ZM7uxudAc} (2022).

\end{thebibliography}






\end{document}